\documentstyle[12pt]{article}
\newcommand{\II}{{\cal I}}
\newcommand{\MM}{{\cal M}}

\newcommand{\be}{\begin{equation}}
\newcommand{\ee}{\end{equation}}
\newcommand{\ben}{\begin{eqnarray}\displaystyle}
\newcommand{\een}{\end{eqnarray}}
\newcommand{\refb}[1]{(\ref{#1})}
\newcommand{\p}{\partial}
\newcommand{\sectiono}[1]{\section{#1}\setcounter{equation}{0}}

\begin{document}

{}~ \hfill\vbox{\hbox{hep-th/9604070}\hbox{MRI-PHY/96-12}}\break

\vskip 3.5cm

\centerline{\large \bf Duality and Orbifolds}

\vspace*{6.0ex}

\centerline{\large \rm Ashoke Sen\footnote{On leave of absence from Tata
Institute of Fundamental Research, Homi Bhabha Road, Bombay 400005, INDIA}
\footnote{E-mail: sen@mri.ernet.in, sen@theory.tifr.res.in}}

\vspace*{1.5ex}

\centerline{\large \it Mehta Research Institute of Mathematics}
 \centerline{\large \it and Mathematical Physics}

\centerline{\large \it 10 Kasturba Gandhi Marg, Allahabad 211002, INDIA}

\vspace*{4.5ex}

\centerline {\bf Abstract}

We construct several examples where duality transformation commutes
with the orbifolding procedure even when the orbifolding group does
not act freely, and there are massless states from the
twisted sector at a generic point in the moduli space. Often the
matching of spectrum in the dual theories
is a result of non-trivial identities satisfied by
the coefficients of one
loop tadpoles in the heterotic, type II and type I string theories.

\vfill \eject

\baselineskip=18pt

\sectiono{Introduction}

Over the last few years many examples of pairs of string theories have been 
constructed such that they are dual to each other in the non-perturbative
sense. In establishing the duality between a pair of theories, one often
uses the power of supersymmetry and the associated non-renormalization
theorems to compare physical quantities which can be calculated in both
the theories. Once duality is established to a resonable level of
confidence, one can then use it to compute physical quantities in
one theory by mapping it to a simpler problem in the dual theory.
This procedure has yielded many
non-perturbative results in string theories which were beyond the scope
of the conventional perturbative formulation of string theory.  

However, in order to be
able to make full use of the power of duality, one should have an {\it
a priori} means of determining when two theories might be dual to each
other. This is particularly relevant in case of pairs of theories with
little or no supersymmetry, where the usual non-renormalization theorems
are absent, and hence we do not have the tools for testing the duality
conjecture. Some progress has been made in this direction, and many of
the duality conjectures have been `derived' from other duality
conjectures by a set of well defined operations. This involves 
construction of new pairs of dual theories by taking orbifolds of known pairs
or a slight variant of this idea\cite{FHSV,VAFAWIT}.
More specifically, suppose theory $A$ compactified on a manifold $K_A$ is
known to be dual to the theory $B$ compactified on the manifold $K_B$.
Let us now compactify both theories further on a manifold $M$. 
Let $G_A$ be a group of discrete symmetries acting on the first theory,
and $G_B$ be the image of this group in the second theory. 
Then naively one would expect
that the first theory modded out by the group $G_A$ will be dual to the
second theory modded out by the group $G_B$. This procedure yields
correctly a pair of dual theories when the duality transformation 
relating the two theories is part of the $T$-duality group, but it does
not always work when the duality is part of the more general $U$-duality
group\cite{VAFAWIT}. Still, it has quite often lead correctly to a
pair of dual theories. The purpose of this paper is to investigate 
this procedure in some detail through some examples.

For simplicity
of argument let us restrict our discussion
to the case where $G_A$ ($G_B$) is a $Z_2$
group generated by the element $g_A$ ($g_B$). Then $G_A=\{1, g_A\}$ and
$G_B=\{1,g_B\}$. By an abuse of notation we shall denote the group
$G_A$ ($G_B$) by its generator $g_A$ ($g_B$), and denote the theory 
$A$ ($B$) on $M\times K_A$ ($M\times K_B$) modded out by $G_A$ ($G_B$)
as theory $A$ on $(M\times K_A)/g_A$ (theory $B$ on $(M\times K_B)/g_B$). 
Let $h_A$ ($h_B$) denote the part of the $Z_2$
transformation representing geometric action on the 
internal manifold $K_A$ ($K_B$) as well as any internal symmetry
transformation and $s$ be the part of the transformation representing
geometric action on the
manifold $M$. Then $g_A=s\cdot h_A$ and $g_B= s\cdot h_B$. We shall now
consider three separate cases:

\begin{enumerate}

\item{ Suppose $s$ acts freely on $M$. In this case the 
manifold $(M\times K_A)/s\cdot h_A$ 
can be regarded as a fiber bundle with base 
space $M/s$ and fiber $K_A$, and similarly the 
manifold $(M\times K_B)/s\cdot h_B$ 
can be regarded as a fiber bundle with base 
space $M/s$ and fiber $K_B$. $h_A$ ($h_B$) represents twist on the
fiber as we move along the order two closed cycles of $M/s$ connecting the
points $P$ and $s(P)$ on $M$. 
The duality transformation 
between the two theories can be regarded as a duality between the
fibers at every point in $M/s$. 
We expect this duality to be valid when the volume of $M$ is
large, since locally 
$M/s$ is nearly flat  and hence the required duality reduces to
the original duality conjecture between theory $A$ on $K_A$ and theory
$B$ on $K_B$. We can then adibatically reduce the volume of $M$ to any
desired size without destroying the duality in this process. This 
argument, due to Vafa and Witten\cite{VAFAWIT} has been named adiabatic
argument.}

\item{Next we consider the case where $s$ does not act freely on $M$, 
but it does not leave the whole of $M$ invariant. In other words $s$ has
non-trivial action on $M$ and has isolated fixed points (or fixed 
hyper-planes). In this case the manifold $(M\times K_A) /s\cdot h_A$ 
can still be regarded as a fiber bundle with base 
space $M/s$ and fiber $K_A$ but the description breaks down at the
orbifold points of $M/s$. At these points the fiber is 
$K_A/h_A$. A similar description holds for the other theory. Thus
we can again try to establish the duality between the two theories by
taking $M$ to be of large volume and applying the duality fiberwise. But
this breaks down at the orbifold points and hence the argument is not as
strong as in the previous case. Nevertheless, this procedure (or a variation
of the procedure) has yielded many correct pairs of dual theories. Let us
now subdivide these models into two classes:
\begin{enumerate}
\item{Suppose that even though $s$ does not act freely on $M$, $s\cdot h_A$
($s\cdot h_B$) acts freely on $M\times K_A$ ($M\times K_B$). In other
words $h_A$ ($h_B$) acts freely on $K_A$ ($K_B$) and hence the fibers at
the orbifold points of $M/s$ are non-singular. Among other things,
this implies that at a generic point in the moduli space (say where
$M\times K_A$  and $M\times K_B$ have large volume) there are no 
massless states from the twisted sector. Thus the spectrum of massless 
states in the two theories after orbifolding match trivially, since
they come from the untwisted sector. There could be special points in the
moduli space where twisted sector states become massless, and comparing 
the effects of these massless states in the two theories provide a
non-trivial check on the duality conjecture. 
The first such construction was given in ref.\cite{FHSV}.}
Most of the examples of
dual pairs, constructed by the orbifolding procedure, fall either 
in this class, or in the previous class where the adiabatic argument is
applicable.
\item{We can also consider the case where $h_A$ ($h_B$) does not act
freely on $K_A$ ($K_B$). In this case $s\cdot h_A$ ($s\cdot h_B$)
does not act freely on $M\times K_A$ ($M\times K_B$),
and often we have massless states from the twisted sector at a generic point
in the moduli space. Thus a non-trivial test of the duality between the
two resulting theories is provided by comparing the spectrum of massless
states from the twisted sectors in the two theories. Most of our
analysis in this paper will focus on this class of theories. We shall 
construct several examples of dual pairs where on each side there are 
massless states from the `twisted sector', and verify 
that the spectrum of these massless states in the two theories agree.}
\end{enumerate}
}
\item{Finally consider the case where $s$ acts trivially on $M$, {\it
i.e. } it leaves the whole of $M$ invariant. In this case 
the manifold $M\times K_A/s\cdot h_A$ does not have the
structure of a fiber bundle with fiber $K_A$ even locally. In fact the
fiber is everywhere $K_A/h_A$. Thus we cannot hope to establish
the duality between the two resulting theories by applying the original
duality on the fibers. Thus the case for equivalence between the two
resulting theories is weakest in this case, and indeed, as we shall
discuss, in most examples of this kind the duality does not hold.}

\end{enumerate}

In the next section we shall consider several examples of dual pairs of
the kind 2(b) and show that the spectrum  of massless  states from the
twisted sector matches in the two theories constructed this way.
Already examples of dual pairs of this kind were constructed in 
ref.\cite{MORBI} (see also \cite{ACHA})
where one side of the theory involved $M$-theory
orbifolds and were shown to reproduce the conjectures of 
refs.\cite{MUKHI,WITTK3}. In this paper we shall focus on examples where both
sides involve orbifolds of string theories, so that one can
independently compute the spectrum of `twisted sector states' in
both theories and make a meaningful comparison. As we shall see, in
several examples, part of the massless spectrum arises from the 
background elementary strings which need to be introduced in the theory
to cancel one loop tadpoles\cite{VWIT}, and only after taking into
account the fields living on these elementary strings, the massless
spectrum in the two theories agree.

\sectiono{Examples of Dual Pairs}

In this section we shall construct examples of dual pairs of theories
using the procedure outlined in the introduction. First
we shall introduce some notations for various symmetry transformations
that we shall be using in our analysis. In type IIA or IIB string
theories, we shall denote by $F_L$ the space-time fermion number
arising in the left-moving sector on the world sheet. Both these
theories are invariant under the $Z_2$ group of transformations
generated by $(-1)^{F_L}$, whose effect on the bosonic fields in the
theory is to change the sign of all
the fields arising in the Ramond-Ramond
(RR) sector, leaving the fields from the Neveu-Schwarz
Neveu-Schwarz (NS) sector invariant. Type IIB theory is also
invariant under
the world-sheet parity transformation which we shall denote by
$\Omega$. Acting on the massless bosonic sector states in the theory,
it changes the sign of the anti-symmetric tensor field in the NS sector,
and the scalar and the rank four anti-symmetric tensor field in the RR
sector, leaving the other fields invariant. Any string theory
compactified on a $2n$ dimensional torus $T^{2n}$ is invariant under
a change of sign of all the $2n$ coordinates on the torus. 
We shall denote this transformation by $\II_{2n}$.
For heterotic string theory compactified
on $T^{2n}$, we shall also define the transformation $\II_{2n+16,2n}$
to be the one that changes the sign of all the coordinates of the 
signature $(2n+16,2n)$ Narain lattice. 

The non-perturbative duality symmetries that we shall be using for
constructing dual pairs of theories are the following. First of all,
in ten dimensions, type IIB theory has an SL(2,Z) $S$-duality 
symmetry\cite{HT}. We shall denote by $S$ the non-trivial $Z_2$ 
transformation
that takes the string coupling to its inverse for vanishing axion field.
For type IIB or IIA theory compactified on $T^4$, the full $U$-duality
group $SO(5,5;Z)$ contains a  $Z_2$ 
transformation that changes the sign of the dilaton field, and takes
the field strength associated with the rank two antisymmetric tensor
field in the NS sector to its dual\cite{SENVAFA}. We shall denote this
by $\sigma$. Finally we shall denote by $\eta$ the string-string
duality transformation
that relates the type IIA string theory compactified on
$K3$ to heterotic string theory compactified on $T^4$\cite{HT}.

Before we discuss construction of dual pairs of type 2(b) mentioned
in the introduction,
let us discuss some examples of type 3 which do not work.
\begin{itemize}
\item{ Consider
first type IIB theory in ten dimensions. In this case it can be easily
seen that conjugation by $S$ takes
$(-1)^{F_L}$ to $\Omega$. Now, type IIB theory
modded out by $(-1)^{F_L}$ gives type IIA theory, whereas the same
theory modded out by $\Omega$ gives the
type I string theory with gauge group SO(32).
These two theories are clearly not the same.}
\item{
For type IIA theory on $T^4$, the transformation $\sigma$ converts
$(-1)^{F_L}$ to $\II_4$\cite{SENVAFA}. Type IIA on $T^4$ modded out
by $(-1)^{F_L}$ gives type IIB on $T^4$, whereas the same theory
modded out by $\II_4$ gives type IIA on $K3$ orbifold. Again the
two theories are clearly not equivalent.}
\item{
Finally let us consider the string-string duality transformation $\eta$.
This maps the symmetry $(-1)^{F_L}$ on the type IIA side to
$\II_{20,4}$ on the heterotic side\cite{VAFAWIT}. Modding out 
by $(-1)^{F_L}$
gives type IIB on $K3$ on the type II side; whereas modding
out by $\II_{20,4}$ gives an
inconsistent theory on the heterotic side because of problem with
left-right level matching\cite{VAFAWIT}.}
\end{itemize}

Note that in each of the cases discussed above we did not compactify
the original model further on another manifold $M$ and combined the
original $Z_2$ symmetry with a non-trivial action on this manifold.
Thus these examples all belong to class 3 where the argument for
duality is the weakest.
We shall now show that in each of the above cases, when
we combine the transformations discussed above with a $Z_2$ action on
the rest of the manifold  we can get sensible
dual pairs. The case where the $Z_2$ action on the rest of the manifold
is free has already been discussed elsewhere\cite{VAFAWIT,SENVAFA},
so we shall focus our attention to the cases where the $Z_2$ action on
the manifold $M$ is not free.

\subsection{Type IIB on  $T^4/(-1)^{F_L}\cdot\II_4$ and type IIB on
$T^4/\Omega\cdot\II_4$}

The ten dimensional $S$-duality transformation $S$ relates the
symmetries $(-1)^{F_L}$ and $\Omega$ in the type IIB theory. We 
compactify the theory on $T^4$ and combine this internal symmetry
with the reflection $\II_4$ of $T^4$. This leads to the dual pair
type IIB on 
$T^4/(-1)^{F_L}\cdot\II_4$ and type IIB on
$T^4/\Omega\cdot\II_4$. An $R\to (1/R)$ duality transformation on one of
the circles of $T^4$ converts the type IIB theory to type IIA theory,
and the transformation $(-1)^{F_L}\cdot \II_4$ to $\II_4$.
Thus the  first theory is related by a $T$-duality transformation to
type IIA on $T^4/\II_4$, which is just the type IIA theory compactified
on a $K3$ orbifold. On the other hand, making a $T$-duality transformation
on all the four circles of $T^4$, the symmetry $\Omega\cdot\II_4$ can be
mapped to $\Omega$\cite{ORIENT}. 
Thus the second theory is related by a $T$-duality
transformation to type IIB on $T^4/\Omega$, which is just the type I
theory on $T^4$. Thus the dual pair constructed here is related by 
$T$-duality to the well known dual pair, type IIA on $K3$ and type I on
$T^4$ (which in turn is known to be equivalent to heterotic string
theory on $T^4$\cite{WIS}).

Using this prescription one can in fact derive a precise
map between the moduli
spaces of the two theories. The particularly interesting aspect of this
is the map between the moduli coming from the `twisted sector' states.
For this it is best to work with the original dual pair instead of their
$T$-dual versions. For the type IIB on $T^4/(-1)^{F_L}\cdot\II_4$, it has
been argued by Kutasov\cite{KUTASOV} (see also \cite{BSV,UNP})
that the twisted sector states
live on the sixteen NS five-branes of the type IIB theory,
moving on $T^4/\II_4$. Each such five-brane supports one vector multiplet
of the non-chiral $N=2$ supersymmetry algebra in six dimensions. The 
vector-multiplet moduli associated with the blowing up modes
in the type IIA description correspond in the type IIB description
to the locations of these 
five-branes on $T^4/\II_4$. On the other hand, for type IIB on 
$T^4/\Omega\cdot\II_4$, the `twisted sector' states live on sixteen
RR five-branes moving on $T^4/\II_4$\cite{ORIENT}. 
Again each of these five-branes support
one vector multiplet of the $N=2$ supersymmetry algebra,
and the vector multiplet moduli from the `twisted sector' correspond to
the locations of the RR five-branes on $T^4/\II_4$. The $S$-duality
transformation in the ten dimensional type IIB theory precisely
transforms an NS five-brane to an RR five-brane and vice versa. This
gives us a precise map between the moduli fields in the two theories:
the moduli coming from the untwisted sectors get mapped into each other
by the usual rules of $S$-duality transformation in the ten dimensional
type IIB theory, and the locations of the
NS five-branes in one theory get mapped to the locations of the RR
five-branes in the dual theory. This story, already anticipated in
refs.\cite{KUTASOV,BSV,KUMAR} in this case, will repeat itself in every 
example that we shall consider below.

\subsection{Type IIB on  $T^8/(-1)^{F_L}\cdot\II_8$ and type IIB on
$T^8/\Omega\cdot\II_8$}

In order to construct this dual pair,
we compactify type IIB theory on an eight dimensional
torus $T^8$ and repeat the procedure of the last subsection with
$\II_4$ replaced by $\II_8$. By an
$R\to (1/R)$ duality transformation in one of the circles of
$T^8$, the first theory is mapped to type IIA on $T^8/\II_8$.
On the other hand by making $R\to (1/R)$ duality transformation on
all the circles, the second theory is mapped to type I theory on
$T^8$. At a generic point
in the moduli space, the `twisted sector states' in the second theory
correspond to sixteen vector multiplets\footnote{By an abuse of notation
we shall refer to the supermultiplet, obtained by dimensional reduction
of the vector multiplet of the N=4 supersymmetry algebra in four 
dimensions, as the vector multiplet of the two dimensional supersymmetry
algebra.} 
of the non-chiral $N=16$ 
\big((8,8)\big)
supersymmetry algebra in two dimensions. Viewed as the 
orientifold IIB on $T^8/\Omega\cdot\II_8$, the `twisted sector' states
live on the sixteen RR one-branes moving on $T^8/\II_8$, and
the 16$\times$8 scalars
from the sixteen vector multiplets may be regarded as the locations
of these sixteen one-branes on $T^8/\II_8$. 

Let us now turn to the twisted sector states in the first theory by
regarding it as type IIA on $T^8/\II_8$. First
of all it is easy to see that there are no massless states from the
sectors where either the left or the right moving fermions are in
the NS sector, since this gives the total vacuum energy in the 
corresponding
sector to be (1/2). Thus the only possible massless states could come
from the RR sector. To determine this spectrum let us work in the light
cone gauge NSR formalism.\footnote{Since light cone gauge analysis
gives us the spectrum for $k^+\ne 0$ states only, working in this gauge
can sometimes be deceptive for chiral theories in two dimensions. Since
the theory we are analysing at present is non-chiral, working in the light
cone gauge does not cause any problem.}
In the twisted RR sector, all the world-sheet
fermions and bosons are anti-periodic, and hence, before the GSO
projection, there is a unique $\II_8$ invariant
ground state with zero energy associated
with each of the 256 fixed points. The transformation $\II_8$,
acting on the RR ground state, has the same effect as $(-1)^{f_L}(-1)^{f_R}$,
where $f_L$ and $f_R$ are the {\it 
world sheet fermion number operators} on the left and the
right moving sectors respectively. We shall work in the convention
where GSO projection requires a Ramond sector state in the left to
have $(-1)^{f_L}$ eigenvalue $-1$ and a Ramond sector state on the
right to have 
$(-1)^{f_R}$ eigenvalue $+1$, reflecting the choice of opposite space-time
chiralities in the left and the right sector in the type IIA theory.
This shows that all the $\II_8$ invariant RR ground states are
odd under either the left or the right moving GSO operator and hence 
are projected 
out. \footnote{When viewed as compactification of type IIB theory
on $T^8/(-1)^{F_L}\cdot\II_8$, GSO projection will require both
$(-1)^{f_L}$ and $(-1)^{f_R}$ eigenvalues to be $+1$. But acting on
RR ground states $(-1)^{F_L}\cdot\II_8$ will have the effect of
$-(-1)^{f_L}(-1)^{f_R}$. Thus $(-1)^{F_L}\cdot\II_8$ invariant states
still fail to survive the GSO projection.}
Thus there are no physical massless states from the twisted sector. 
(For definiteness
we shall take the twisted sector RR states to be even under both
$(-1)^{f_L}$ and $(-1)^{f_R}$. Thus they survive the GSO projection on
the right, but fails to survive the GSO projection on the left.)

This seems to lead to a contradiction, since duality implies that this
theory must have sixteen vector multiplets from the twisted sector. 
There is however a subtle effect which restores these vector multiplets
in this theory. As was shown by Vafa and Witten\cite{VWIT}, some string
theories in two dimensions can have tadpoles of the NS sector
anti-symmetric tensor field of the form:
\be \label{e1}
2\pi i n B\, ,
\ee
where
\be \label{e2}
B = {1\over 2} \epsilon^{\mu\nu} B_{\mu\nu}
\ee
is normalized so as to have periodicity one, and $n$ is a constant. The
theory is inconsistent in the presence of such a tadpole, but this
can be overcome by placing $n$ elementary strings moving in the internal 
manifold, whose word-volume fills up the physical space-time. Each of
these strings would give a vector multiplet of the $N=16$ supersymmetry
algebra in two dimensions, with its eight scalar components labelling
the location of the string in the internal space. Thus if in the present
case $|n|$ turns out to be 16, then we would get complete agreement 
between the spectrum of massless states in the two theories. This
would also give a map between the moduli spaces 
in the two theories as in the previous case.\footnote{In a two
dimensional theory the word moduli space is a misnomer, but we shall
continue to use it anyway to denote the configuration
space of massless scalars in the theory.} To see this we work 
with the original pair, type IIB on $T^8/(-1)^{F_L}\cdot \II_8$ and
type IIB on $T^8/\Omega\cdot \II_8$. The moduli from the untwisted
sector states in the two theories are mapped into each other by
the usual $S$-duality transformation rules of the ten dimensional
theory. The moduli of the twisted sector states in the first theory
are represented by the locations of the sixteen elementary type IIB
strings on $T^8/\II_8$ (note that under the $R\to 1/R$ duality that
takes us from the type IIB to the type IIA description, the elementary
type IIB string goes over to the elementary type IIA string). On the
other hand the moduli of the twisted sector states of the second theory
are represented by the locations of the sixteen RR strings (D-strings)
on the internal manifold $T^8/\II_8$. The $S$-duality transforms
the elementary type IIB string to a D-string, and hence the locations of
the elementary strings in the first theory get mapped to those of the
$D$-strings in the second theory. 

Thus all that remains to be shown is that the number $|n|$ of elementary
strings, required to cancel the tadpole of the $B$ field, is indeed 16
in the present case. This number was calculated in 
ref.\cite{VWIT} and is given by (up to a sign):
\be \label{e3}
n = -{1\over 8\pi} \int_{\p\MM} d\tau_1 \Big({1\over \tau_2}
- {4i\p_\tau\eta \over \eta}\Big) A_M(q)\, 
\ee
where $\MM$ denotes a fundamental region of the moduli space of
a torus, $\tau=\tau_1+i\tau_2$ is the complex coordinate labelling
this moduli space, $\eta$ is the Dedekind eta function, and $A_M(q)$
is the elliptic genus of the conformal field theory associated with
the eight transverse coordinates, defined as the partition function
of this conformal field theory in the (even,odd) spin structure
where even and odd refer to the left and the right moving sectors on
the world-sheet. Defining $q=e^{2\pi i\tau}$, and noting that
$\int_{\p\MM} d\tau_1=\ointop (dq /2\pi iq)$ we see that the
non-trivial contribution from the boundary $q=0$ to the above 
integral can be obtained by expanding the integrand in powers of
$q$ and keeping the $q^0$ term in the expansion. 
In particular
\be \label{e4}
{1\over \tau_2}- {4i\p_\tau\eta\over \eta} = {\pi\over 3}
-{2\pi \over \ln|q|} - 8 \pi q + O(q^2)\, .
\ee
On the other hand\cite{VWIT}, 
\be \label{e5}
A_M(q) = -(2 n_{NS,R} - n_{R,R}) + O(q)\, ,
\ee
where $n_{NS,R}$ is the number of massless states in the NS-R sector that
survives GSO projection from the left, weighted by $(-1)^{f_R}$ where
$f_R$ denotes the world-sheet fermion number from the right.
Note that in counting $n_{NS,R}$ we do not require the state to have
survived GSO projection on the right. $n_{R,R}$ on the other hand is
the number of massless states in the RR sector weighted by $(-1)^{f_R}$;
in this case we do not require the state to have survived GSO
projection either from the left or from the right. Using \refb{e4}
and \refb{e5} in \refb{e3} we get
\be \label{e6}
n = {1\over 24} (2 n_{NS,R} - n_{R,R})\, .
\ee

Thus it remains to compute $n_{NS,R}$ and $n_{R,R}$. First we shall
compute $n_{R,R}$. This can get contribution from the untwisted
sector as well as the twisted sector. Before GSO projection, and
the projection by $\II_8$, there are $16\times 16$ states from the
untwisted sector, which transform under the tangent space 
$SO(8)$ group of the internal manifold as 
\be \label{e7}
(8_s+8_c)_L \times (8_s+8_c)_R\, ,
\ee
$8_s$ and $8_c$ denoting the two inequivalent spinor representations
of $SO(8)$. $\II_8$ acts on these spinor representations of $SO(8)$
by changing the sign of $8_s$ but keeping $8_c$ invariant. Thus the
states that survive the $\II_8$ projection are
\be \label{e8}
(8_c)_L\times (8_c)_R + (8_s)_L\times (8_s)_R\, .
\ee
Since we have adopted the convention that the action of $\II_8$ and
$(-1)^{f_R}$ agree on the Ramond sector ground states on the right,
$(-1)^{f_R}$ leaves $(8_c)_R$ invariant and changes
the sign of $(8_s)_R$. Thus of the 128 states that survive the $\II_8$
projection, 64 are odd under $(-1)^{f_R}$ and 64 are even. As a result
the contribution of these states to $n_{R,R}$, which counts the 
number of massless RR states weighted by $(-1)^{f_R}$, vanishes.

Let us now turn to the twisted sector. As mentioned before, there
are 256 fixed points, and associated with each of them is a
unique ground state of zero energy in the RR sector that is
invariant under $\II_8$ (but does not survive the GSO projection on
the left). According to the convention we have adopted,
these all carry $(-1)^{f_R}=1$. This gives
\be \label{e9}
n_{R,R}=256\, .
\ee

Computation of $n_{NS,R}$ follows a similar pattern. The untwisted
sector states in the massless sector before any projection transform
under SO(8) as
\be \label{e10}
(8_v)_L \times (8_s+8_c)_R\, ,
\ee
where $8_v$ denotes the vector representation of SO(8).
This survives the GSO projection on the left. Under $\II_8$,
$8_v$ and $8_s$ changes sign, but $8_c$ remains invariant. Thus
the state that survives $\II_8$ projection is $(8_v)_L
\times (8_s)_R$. Since $(8_s)_R$ has $(-1)^{f_R}=-1$, these 
states contribute a net factor of $-64$ to $n_{NS,R}$.
In the twisted sector, all the eight bosons on the left
are antiperiodic and all the eight fermions on the left are periodic. This
gives a total contribution of (1/2) to the vacuum energy on the
left, showing that there are no massless states from this sector.
Thus we get 
\be \label{e11}
n_{NS,R}=-64\, .
\ee
Using eqs.\refb{e6}, \refb{e9} and \refb{e11} we finally get
\be \label{e12}
|n| = 16\, ,
\ee
as required.

\subsection{Type IIA on  $(T^4 \times (T^4)')/(-1)^{F_L}\cdot\II_4'$ 
and type IIA on $(T^4 \times (T^4)')/\II_4\cdot\II_4'$}

Type IIA (or IIB) theory compactified on $T^4$ has a $Z_2$ self-duality
transformation $\sigma$ described before. This converts the 
transformation $(-1)^{F_L}$ to $\II_4$. Let us compactify the theory
further on a four torus (which we shall denote by $(T^4)'$) and mod
out the theory by combined operation of $(-1)^{F_L}$ (or $\II_4$)
and the reflection $\II_4'$ on this new torus. This leads to the
dual pair of theories described in the title of this subsection.

We shall now compare the spectrum of twisted sector states in the two
theories obtained this way. The first theory, after an $R\to 1/R$ 
duality transformation in one of the circles of $(T^4)'$ can be
identified to type IIB on $(T^4\times (T^4)')/\II_4'$, {\it i.e.}
type IIB on $T^4\times K3'$ at the orbifold limit of $K3'$. The
twisted sector states in this case come from the sixteen fixed points 
of $(T^4)'$ under $\II_4$. Following Kutasov\cite{KUTASOV}
it is more convenient to describe these in the original orbifold $-$
type IIA on $(T^4\times (T^4)')/(-1)^{F_L}\cdot\II_4'$ $-$ as living on
sixteen NS five-branes of the type IIA theory transverse to $(T^4)'/\II_4'$
and wrapped around $T^4$. In type IIA on $T^4$ such a
configuration corresponds to the
dual of the elementary string in six dimensions. Thus the twisted
sector states can be regarded as living on the sixteen dual strings
in six dimensions moving on $(T^4)'/\II_4'$.

In the second theory, the transformation $\II_4\times \II_4'$ is
simply the reflection $\II_8$ of the full eight dimensional torus. 
Thus this theory is just type IIA on $T^8/\II_8$, $-$ the one we
analysed in the previous subsection. As was found there, in this case
there are no massless states from the conventional twisted sector, but
cancellation of one loop tadpoles forces us to introduce sixteen
elementary strings. There are extra massless
states living on these sixteen elementary strings
moving on $T^8/\II_8$. An elementary string in the ten dimensional
type IIA string is also an elementary string in the six dimensional
type IIA string obtained by compactifying the ten dimensional theory on
$T^4$. Thus from the six dimensional view point, the `twisted sector
states' live on sixteen elementary strings moving on $(T^4)'/\II_4'$.

Comparison with the first theory shows that this is precisely what we
would have expected to happen. Under the duality transformation $\sigma$
in the six dimensional theory,
dual strings get converted to elementary strings
and vice versa. This converts the sixteen dual strings in the first
theory to sixteen elementary strings in the second theory. Besides
showing that the spectrum in the two theories agree, this procedure again
gives us the relation between the moduli fields in the two 
theories, in the untwisted as well as the twisted sector.

\subsection{Type IIA on  $(K3 \times (T^4)')/(-1)^{F_L}\cdot\II_4'$ 
and heterotic theory on $(T^4 \times (T^4)')/\II_{20,4}\cdot\II_4'$}

Type IIA theory compactified on $K3$ is expected to be dual to the
heterotic string theory on $T^4$. As was shown in ref.\cite{VAFAWIT},
under this duality, the
transformation $(-1)^{F_L}$ in the type IIA theory gets mapped
to $\II_{20,4}$ in the heterotic theory, where $\II_{20,4}$ denotes the
change of sign of all the coordinates on the signature (20,4) 
Narain lattice. Let us compactify the theory
on a further four torus (which we shall denote by $(T^4)'$) and mod
out the theories by combined operation of $(-1)^{F_L}$ (or $\II_{20,4}$)
and the reflection $\II_4'$ on this new torus. This leads to the
dual pair of theories described in the title of this subsection.

The spectrum in the twisted sector of the first theory can be analysed
as in the previous example. By an $R\to (1/R)$ duality transformation
in one of the circles of $(T^4)'$ this theory gets mapped to type IIB
on $(K3\times (T^4)')/\II_4'$, {\it i.e.} type IIB on $K3\times K3'$
in the orbifold limit of $K3'$. The twisted sector states come from the
sixteen fixed points of $(T^4)'$ under $\II_4'$. In the original version
of the theory before the $T$-duality transformation, these states
could be interpreted as living on the sixteen NS five-branes of the
type IIA theory transverse
to $(T^4)'$, and wrapped around $K3$. In the six dimensional
language, this represents sixteen dual strings moving on $(T^4)'/\II_4'$.
Note that this theory does not have any tadpole of the $B_{\mu\nu}$
field\cite{VWIT}, thus we do not get any extra elementary type II strings
in this theory (unlike in the case of type IIA theory on $K3\times K3$).

Let us now turn to the heterotic side of the story. The transformation
$\II_{20,4}\times \II_4'$ represents the change of sign of all the 
directions of the signature $(24,8)$ Narain lattice representing heterotic
string compactification on $T^8$. Let us denote this transformation
by $\II_{24,8}$. Modding out heterotic string theory on $T^8$ by
$\II_{24,8}$ gives a consistent modular invariant theory, since 24 is
a multiple of 8 and hence there is no problem with level matching 
between the left and the right sectors. Examining the twisted sector
states, however, we discover that there are no massless states
from the twisted sector. This is due to the fact that twenty four
left-moving twisted bosons contribute a total vacuum energy of $1/2$.
This seems to
lead to an apparent contradiction, since the orbifold of the type IIA theory 
has massless states from the twisted sector. The resolution to
this comes from the fact that the heterotic theory again has a one loop
tadpole of $B_{\mu\nu}$ field of the form \refb{e1}, which forces us to
introduce $|n|$ elementary heterotic strings as background. In hindsight,
this is precisely what we should have expected. On the type IIA side the
twisted sector states live on sixteen dual strings, which under 
string-string duality transformation get mapped to sixteen 
elementary heterotic strings. Thus in order to show that the spectrum
of massless states in these two theories coincide, all we need to show
is that in this case $|n|$ equals 16.

The computation of $n$ in the heterotic string theory proceeds in the
same way as in the type IIA case and the final formula for $n$ is
identical to the one given in eq.\refb{e3}. There are two factors of
2 compared to the type II calculation  which cancel. First of all
GSO projection gives only a factor of (1/2) in this case instead of a
factor of (1/4), since we have GSO projection only on the right.
Also in this case spin structure arises only in the right sector, and
unlike in the type II case, where (odd,even) and (even,odd) spin
structures give equal contribution giving an extra factor of 2, here
the only contribution comes from the odd spin structure on the right
without giving this extra factor.

Thus in order to calculate $n$, we need to calculate
$A_M(q)$ for the conformal field theory describing the dynamics of
transverse and internal coordinates of the 
heteroic string. This receives contribution from the
Ramond sector in the right. As in the type IIA case we shall expand
$A_M(q)$ in powers of $q$ and keep only terms upto order $q^0$. 
The twisted sector does not contribute, since the total vacuum energy of
twenty four left-moving bosons exceeds zero. Untwisted sector contribution
is given by:
\be \label{e13}
A_M(q)=-8 \big(q^{-1}-24 + O(q)\big)\, .
\ee
This arises in the following way. First of all the overall minus sign
in $A_M(q)$ is due to the fact that the trace in this case is taken over
the space-time fermionic states. There are sixteen
Ramond sector ground states on the right, which transform in the $(8_s+8_c)$
representation of SO(8). Of these $8_s$ is odd and $8_c$ is even under
$(-1)^{f_R}$ as before. Also $8_s$ is odd and $8_c$ is even under 
$\II_{24,8}$. The unique ground state on the left is even under 
$\II_{24,8}$ and hence must be combined with the state $8_c$ on the right
to give an $\II_{24,8}$ invariant state. These eight states have
vacuum energy of $-1$
from the left, and are even under $(-1)^{f_R}$; hence they give a 
contribution
of $8q^{-1}$. The first excited state on the left is twenty four fold
degenerate and is odd under $\II_{24,8}$. 
Thus it must be combined with $8_s$
on the right to give an $\II_{24,8}$ invariant state. 
These 24$\times8$ states
have vanishing vacuum energy from the left sector, and are odd under
$(-1)^{f_R}$; hence they contribute $-8\cdot24$. 

Substituting \refb{e13} and
\refb{e4} into \refb{e3} we get
\be \label{e14}
|n|=16\, ,
\ee
as is required for getting duality invariant spectrum of massless states.

\subsection{Type IIA on  $(K3 \times 
(T^4)')/(-1)^{F_L}\cdot\sigma_{II}\cdot\II_4'$ 
and heterotic theory on $(T^4 \times (T^4)')/\II_{20,4}\cdot\sigma_H
\cdot\II_4'$}

We consider type IIA theory compactified on a special class of $K3$
surfaces which have a $Z_2$ isometry generated by $\sigma_{II}$ with the
following properties\cite{NIKULIN,SCHSEN,CHLO}:
\begin{enumerate}
\item{It exchanges the two $E_8$ factors in the lattice of second cohomology
elements of $K3$.}
\item{It has eight fixed points on $K3$.}
\item{Modding out by this symmetry gives us back an orbifold of SU(2) 
holonomy.}
\end{enumerate}
The corresponding transformation $\sigma_H$
in the dual heterotic string theory on
$T^4$ simply exchanges the two $E_8$ gauge groups in the theory.
We now further compactify both theories on a four torus $(T^4)'$ and
mod out the type IIA theory by $(-1)^{F_L}\cdot\sigma_{II}\cdot\II_4'$
and the heterotic theory by its image $\II_{20,4}\cdot\sigma_H\cdot\II_4'$.
This leads us to the dual pair described above.

We shall now compare the spectrum of massless states in the twisted 
sector in the two theories. 
By making an $R\to(1/R)$
duality transformation on one of the circles of $(T^4)'$ we can map the
type IIA 
theory to type IIB on $(K3\times(T^4)')/\sigma_{II}\cdot\II_4'$.
In this theory there are $8\times16$
fixed points, since $\sigma_{II}$ has eight fixed points on $K3$ and
$\II_4'$ has sixteen fixed points on $(T^4)'$. 
The twisted sector 
at each of these fixed points is characterized by eight anti-periodic
bosons on the left and eight anti-periodic bosons on the right. 
The eight fermions on either
side are periodic in the NS sector and anti-periodic in the R sector.
As a result the total ground state energy in either side is zero
in the R sector and one in the NS sector. Thus the only massless states
in the twisted sector arise from the ground state of the RR sector. This
state is unique since all the fermions are anti-periodic. Furthermore,
this state, as constructed, is chiral, since in the light cone gauge that
we are using the $k^+$ component of the momentum is constrained to be
non-vanishing. It is easy to verify by working in the covariant formulation
that there are no massless states 
in this theory from the twisted sector with non-vanishing $k^-$. Thus
each fixed point gives a massless chiral boson, giving a total of 
128 massless chiral bosons in this theory. One can also verify that
there are no $B_{\mu\nu}$ tadpoles in this theory since the type IIB
theory is invariant under a world-sheet parity transformation under
which $B_{\mu\nu}$ changes sign\cite{VWIT}. Thus we do not need to 
introduce any background elementary type II string.

Let us now turn to the twisted sector states in the heterotic theory.
There are altogether $16\times 16$ fixed points since $\II_{20,4}$
has sixteen fixed points on $T^4$ and $\II_4'$ has sixteen fixed points
on $(T^4)'$. The action of the $Z_2$ transformation 
on the sixteen internal left moving bosons is to exchange eight of them
with eight others. Thus by taking appropriate linear combinations of these
bosons we get eight periodic and eight anti-periodic internal bosons in the
twisted sector. On the other hand all the eight bosons associated with the
eight coordinates labelling $T^4\times (T^4)'$ are anti-periodic both 
in the left and the right moving side. Thus we have a total of eight
periodic and sixteen anti-periodic bosons on the left, giving a total
vacuum energy of zero. On the right hand side we have eight anti-periodic
bosons. In the NS sector, the eight fermions on the right
are periodic, giving a total vacuum energy one, whereas in the R sector
the eight fermions on the right are anti-periodic, giving a total 
vacuum energy zero. Thus we get a unique massless state from the Ramond 
ground state on the right. This is a fermionic state, and as before one
can verify by working in the covariant formalism that these are
chiral fermions. Thus we get a total of 256 chiral fermions from the
256 fixed points. Using Bose-Fermi equivalence in two dimensions these
can be shown to be equivalent to 128 chiral bosons. Thus we again get
identical spectrum of massless fields from the twisted sector 
of the two theories.

It remains to verify that there is no $B_{\mu\nu}$ tadpole in this
theory, since any such tadpole will force us to introduce heterotic
string background and hence introduce new massless states from the
collective coordinates of these strings. The calculation proceeds
as in the last subsection. In particular, the untwisted sector
contribution to $A_M(q)$ is now given by
\be \label{e15}
-8(q^{-1}-8+O(q))\, .
\ee
Note that the 24 in eq.\refb{e13} has been replaced by 8, since
of the 24 left moving bosonic oscillators 16 are odd and 8 are even under
$\II_{20,4}\cdot\sigma_H\cdot\II_4$. Thus 16 of these oscillator
states need to be accompanied by $(-1)^{f_R}=-1$ states from the
right, and 8 of them have to be accompanied by $(-1)^{f_R}=1$ states
from the right. This time there is also twisted sector contribution
to $A_M(q)$. The 256 massless twisted sector states, each with 
$(-1)^{f_R}=1$, contributes a factor of $-256$ to $A_M(q)$. (The $-1$
is again due to the fact that these states are all space-time
fermions.) This gives
\be \label{e16}
A_M(q)=-8(q^{-1}+24+O(q))\, .
\ee
Substituting this and \refb{e4} into \refb{e3} we get
\be \label{e17}
n=0\, ,
\ee
as desired.

\section{Conclusion}

In this paper we have constructed several examples where the orbifolding
procedure commutes with the duality transformation. We start with a known
dual pair of theories, identify a pair of symmetries in these two theories
that are related by a duality transformation, and mod out both theories by 
their respective symmetries to construct a dual pair. In most cases, if
we do not combine the original pair of symmetries with a space-time
symmetry transformation,
we are lead to an inconsistent result. On the other hand, if
we combine the original pair of symmetries (which could be called internal
symmetries) with a space-time symmetry transformation (with fixed
points in general), and then construct the orbifold,
we get a consistent dual pair. 
In many of these cases, we get identical spectrum of massless states
in the dual pair of theories constructed this way only after introducing
appropriate background fields that cancel the one loop tadpoles in both
the theories. This puts non-trivial constraint on the coefficients of one
loop tadpoles in various theories, and in every
case that has been studied one finds that the  coefficient of the
tadpole is consistent with the predictions of duality.

Many examples of dual pairs of theories, constructed by modding out 
another dual pair by appropriate symmetries, have
been discussed before\cite{FHSV,VAFAWIT,SENVAFA,ALOK}. 
Our examples differ from most of the previous examples
in that in our models, there are massless  states from the `twisted
sector' in both theories {\it at a generic point in the moduli
space.} So far there has been no systematic rule for
determining when orbifolding commutes with duality transformation, 
except in cases where
the adibatic argument of ref.\cite{VAFAWIT} is applicable.
We hope that the results of this paper will provide a step towards
a more systematic understanding of this phenomenon.

Finally the result of this paper boosts our confidence in the 
results of ref.\cite{MORBI} where many of the conjectures involving
orbifolds of $M$-theory were derived using the ansatz that orbifolding
procedure commutes with the duality transformation. The only place 
where this procedure failed was in finding the dual of the $Z_2$ orbifold
of $M$ theory on $S^1$, where it gave the dual theory as type IIB string
theory in ten dimensions instead of the $E_8\times E_8$ heterotic 
string\cite{HOR}.
According to the classification given in this paper, this example falls
in the class 3, where the argument for duality is the weakest, and fails
even in many string theory examples. On the other hand, every other
example discussed in ref.\cite{MORBI}, where this procedure gave 
sensible answer, is of type 2(b)
in our classification. As we saw in this paper, in many string theory
examples of this kind we get sensible answers by assuming that 
orbifolding commutes with the duality group. It is satisfying that even
in the $M$-theory examples we got sensible answers precisely for this
class of models.

I wish to thank K. Dasgupta and S. Mukhi for useful discussions.
I would also like to thank the theoretical high energy physics group at
Rutgers university for hospitality during the course of this work.

\end{document}